\documentclass[12pt,aps,preprint]{JHEP3}
\usepackage{amsmath,amssymb,graphicx,mathrsfs,epsfig}
\usepackage[latin1]{inputenc}
\usepackage{bm}

\newcommand{\tbt}{(T+ \overline T)}

\newcommand{\be}{\begin{equation}}
\newcommand{\ee}{\end{equation}}
\newcommand{\bea}{\begin{eqnarray}}
\newcommand{\eea}{\end{eqnarray}}
\newcommand{\bt}{\overline{T}}
\newcommand{\tn}{T_0}
\newcommand{\btn}{\overline{T}_0}
\title{Models of Modular Inflation and Their Phenomenological Consequences}
\author{Ido Ben-Dayan, Ram Brustein\\
Department of Physics, Ben-Gurion University,
    Beer-Sheva 84105, Israel \\ E-mail: idobd@bgu.ac.il,ramyb@bgu.ac.il}
\author{ Senarath P. de Alwis\\  Department of Physics,  University of Colorado,  Boulder, CO 80309, U.S.A. \\ E-mail: dealwiss@colorado.edu}
\preprint{}
\abstract{
We study models of modular inflation of the form expected to arise from low energy effective actions of superstring theories. We argue on general grounds that the most likely models of modular slow-roll inflation are small field models in which the inflaton moves about a Planck distance from an extremum of the potential. We then focus on models in which the inflaton is the bosonic component of a single (complex) chiral superfield and explain the generic difficulties in designing small field models of modular inflation. We then show that if the K\"ahler potential of the inflaton is logarithmic as in perturbative string theories, then it is not possible to satisfy the slow-roll conditions for any superpotential.
We find that if the corrections to the K\"ahler potential are large enough so it can be approximated by a canonical K\"ahler potential in the vicinity of the extremum, then viable slow-roll inflation is possible. In this case, several parameters have to be tuned to a fraction of a percent.  We give a prescription for designing successful small field supergravity models of inflation when the K\"ahler potential is canonical and calculate the slow-roll parameters from the superpotential parameters.  Our results strengthen the case for models in which the moduli slowly roll about a Planck distance from a relatively high scale extremum that is located in the vicinity of the central region of moduli space where the coupling and compact volume are both of order unity in string units.  Generic models of this class predict a red spectrum of scalar perturbations and negligible spectral index running. They also predict a characteristic suppression of tensor perturbations despite the high scale of inflation. Consequently,  a detection of primordial tensor anisotropies or spectral index running in cosmic microwave background observations in the foreseeable future will rule out this entire class of  modular inflation models.
}
\keywords{Inflation, String theory and Cosmology}
\begin{document}

\section{Introduction}
\label{intro}

Finding viable models of inflation in string theory has been an outstanding problem
for more than two decades. Closed string moduli (CSM) have been marked as candidate inflatons since the very beginning \cite{binetruy}. During the last two decades many other proposals to incorporate inflation into string theory have been put forward as reviewed in \cite{review1,review2,review3,review4}, notably braneworld models with some or many open string moduli as the inflatons. However, it is fair to say that we still do not have a satisfactory understanding of inflationary dynamics within string theory.

It is well known that the CSM have to be stabilized for any form of inflation in another sector of the theory to take place \cite{bs,KMM}. Otherwise their kinetic energy will dominate, and inflation will not be possible. If some of the CSM are not stabilized at the perturbative string level, as is the case in many situations, then they have to be stabilized either by stringy or field theoretic non-perturbative (NP) effects.  In most examples of successful moduli stabilization models that we have encountered, there are in addition to the ``good minima",  additional ``bad extrema" that miss some of the requirements. This happens because the moduli extrimization conditions $V'=0$ typically have many solutions.   The spurious extrema are either  minima, saddles or maxima. Consider a good minimum at $\phi_0$ ($\phi$ stands for the set of moduli). In its vicinity, the moduli superpotential can be expanded in powers of the deviation $(\phi-\phi_0)$. Generically, the superpotential will not be a low order polynomial, and hence its K\"ahler derivative $D_\phi W$ will have several zeros in a limited region in field space. Each such point where $D_\phi W=0$ is an extremum where the value of the potential is negative. We have found previously \cite{bda1,bda2} that the additional extrema are concentrated within a rather small distance in field space from the good ones, and that they arise for generic choices of stringy parameters such as fluxes and vacuum expectation values of complex structure moduli. Thus, each point in the landscape of stringy solutions has its own fine structure: the ``mini-landscape".

In flux compactifications the number of parameters that can be chosen for model building is quite large and many of them can be tuned in small increments by a discrete choice of the fluxes and other parameters. Hence it seems likely that there are some models that will actually support inflationary dynamics. In this paper we aim to determine whether such tuning is possible and reasonable.

In general, inflation can occur in a complicated multi-dimensional space. The background evolution is not necessarily driven by the same inflaton field whose quantum fluctuations create the cosmological perturbations, etc. However, in most simple cases it is possible to identify, at least a-posteriori, a single inflaton. So to get some insight about the expected typical results it is possible to use effective single field potentials. However, one has to keep in mind that more complicated models can, and sometimes do, produce different results.  Single field models of inflation are conveniently classified into two main classes, large field models and small field models. The large field class contains also some hybrid models. In large field models, as their name suggests, the inflaton $\phi$ moves a large distance in (reduced) Planck scale units while inflation takes place $\Delta\phi\gg m_p$ (our conventions are such that $m_p\equiv\frac{1}{8\pi
G_N}=1.2\times 10^{18} GeV$). In small field models the inflaton moves only a relatively short distance $\Delta\phi\lesssim \text{ a few } m_p$.

Large field models are not likely to be realized as models of modular inflation \cite{cosmoduli}. The argument is based on two key points, that the dynamics of moduli in the outer region of moduli space does not support inflation and that duality symmetries relate the various string theories.
The remaining region that cannot be mapped to any perturbative corner of string theory is not very large, perhaps less than a few Planck units. Since the  gravity  and moduli effective action of all perturbative string theories has the same functional form, the action of dualities on the different effective actions has to be represented by  field redefinitions which are allowed by the symmetries of the lagrangian. The duality transformations must therefore act on the space of solutions by mapping one set of solutions onto another set. In the context of weakly coupled heterotic string theory it is well known that the potentials that are generated for the dilaton and the compactification moduli  are steep and consequently have a problem providing enough slow-roll inflation. The dualities  allow the extension of these arguments to all corners of moduli space. This conclusion remains correct also when one includes the contributions of brane instantons and was checked explicitly for various string theories in \cite{cosmoduli}.  In the effective field theories, any large motions in moduli space necessarily brings the moving field to some perturbative region where its potential can no longer support inflation and then inflation ends. We would like to emphasize that quantum gravity considerations \cite{Lyth} cannot be a generic reason for disfavoring large field models as explained in \cite{lindemukhanov}. Rather the reason that small field models emerge as the most likely candidates comes from some specific properties of the potentials. Small field models are the preferred class of models, however, it has been notoriously difficult to realize them as modular inflation models in the perturbative region of moduli space. We explain later some of the difficulties by a general analysis.

The class of inflationary models that we will focus on are small field models in which inflation occurs near a ``flat feature", a maximum or a saddle point with a small curvature (small second derivatives). An interpolating field configuration may extend on this flat feature and if the extent in field space of the feature is larger than a reduced Planck distance, and the curvature near the top is small enough, a ``topological" defect, such as a domain wall, a monopolole, etc., with an inflating core will form \cite{guendelman,vilenkin,linde}. This is a realization of slow roll inflation which is most appropriate to the ``mini landscape". As we have argued, we do not expect large flat regions in field space, however, we can expect extrema, and some of them may be flat enough. Some additional mild constraints seem to be required to ensure that sufficient inflation can be obtained. In particular, quantum fluctuations need to be small enough so that the field can start close enough to the extremum and slowly roll for enough e-folds.

Consider now a flat feature of the kind that we have discussed in the ``mini-landscape". Moduli superpotentials are
exponentials, and therefore the potentials for canonically normalized fields are
typically exponentials of exponentials, which in the outer region of moduli space
are steep. The second derivative has to be small at the extremum to allow
sufficient inflation, however the third derivative is not necessarily small, and if
for some reason the third derivative is small (for example by symmetry), the fourth
derivative is not necessarily small and so on.  In the literature \cite{kinney,lythtop} the
phenomenology of simple topological inflation models in which the potential can be
approximated as a single power of the field for the duration of inflation are
discussed. They produce a red spectrum of scalar perturbations, negligible spectral index running and a very small
amount of tensor perturbations.

In \cite{cosmocenter} it was argued further  that in the central region of moduli space the potential which is suggested by membrane instanton effects has the correct scaling and shape to allow for enough slow-roll inflation, and to produce the correct amplitude of cosmic microwave background (CMB) anisotropies.  Limited knowledge of some generic properties of the induced potential were sufficient to determine the simplest type of consistent inflationary model and its predictions (about a year before the first results from WMAP were released) about the spectrum of cosmic microwave background anisotropies: a red spectrum of scalar perturbations, and negligible amount of tensor perturbations.

In this paper we continue to study these issues in more detail and we improve and strengthen the results. On the more theoretical side we show that if the K\"ahler potential of the inflaton is logarithmic as in perturbative string theories, then it is not possible to satisfy the slow-roll conditions for any superpotential.
We then show that if the corrections to the K\"ahler potential are large enough so it can be approximated by a canonical K\"ahler potential in the vicinity of the extremum, then viable slow-roll inflation is possible.  We give a prescription for designing successful supergravity models of inflation and calculate the slow-roll parameters from the superpotential parameters.  On the more phenomenological side we verify that generic models of this class predict a red spectrum of scalar perturbations, negligible running of the spectral index and a characteristic suppression of tensor perturbations.  We quantify the running of the spectral index and the suppression of tensor perturbations.

The paper is organized as follows. In section~\ref{smallfields} we review the conditions for inflation in small field models when the inflaton starts near an extremum and explain the generic difficulties of constructing them. In section~\ref{topinflation} we prove that the conditions for inflation cannot be satisfied within single complex field SUGRA models with $K=-A\ln \tbt \quad  0< A\leq 3$ for any regular holomorphic superpotential
(The results of this subsection were also independently obtained by \cite{Badziak:2008yg} ).
Next we show how such single complex field models may arise in perturbative string theory.  Later in section~\ref{topinflation} we demonstrate that for $K=T \bt$ it is possible to have viable inflation provided that the superpotential parameters can be tuned to obey certain equations. We also give some numerical examples and predictions for the spectral index $n_S$, running parameter $\alpha$ and tensor-to-scalar ratio $r$ in terms of the superpotential parameters. In section~\ref{GW} we treat hill-top inflation potentials of the type and analyze their gravitational waves (GW) spectra. We show that  the simplest class of models  yields a very small amount of GW and negligible running parameter. Section~\ref{conc} contains a brief summary of our results and some conclusions.

\section{Small field models of modular inflation}
\label{smallfields}

In this section we review the basic definitions of inflationary models and the rationale leading us to propose that small field models are the most likely models of modular inflation. We also discuss why it is that successful small field models are so hard to construct. In the next section we prove that some major changes have to be made to overcome these difficulties.  Some of the material in this section has appeared in the literature in one way or another, however we are not aware of a coherent exposition of the issue.

\subsection{Definitions}

We start by defining for later use the inflation slow-roll parameters (these and other definitions are reviewed in \cite{lythriotto}),
\begin{equation}
\epsilon(\phi)=\frac{m_p^2}{2} \left(\frac{V'}{V}\right)^2,
\end{equation}
\begin{equation}
\eta(\phi)=m_p^2 \left(\frac{V''}{V}\right)
\end{equation}
\begin{equation}
\xi^2(\phi)=m_p^4\left(\frac{V'''V'}{V^2}\right).
\end{equation}

The number of e-folds can be expressed in terms of the slow roll parameter
$\epsilon$
\begin{equation}
N(\phi)=\frac{1}{\sqrt{2}m_p}
\int_\phi^{\phi_{END}}\frac{d\widetilde{\phi}}{\sqrt{\epsilon\left(\widetilde{\phi}\right)}}.
\end{equation}
Inflation ends when $\epsilon(\phi_{END})=1$.

The power spectrum of scalar perturbations is approximately given by
\begin{equation}
P_S=\left[\left(\frac{H}{m_p}\right)^2 \frac{1}{\pi\epsilon}\right]_{k=aH}
\end{equation}
and the power spectrum of tensor (gravitational waves) perturbations is approximately given by
\begin{equation}
P_T=\left[16 \left(\frac{H}{m_p}\right)^2 \frac{1}{\pi}\right]_{k=aH}.
\end{equation}
The expressions are evaluated at horizon exit $k=aH$.
The ratio of tensor to scalar amplitude  $r=\frac{P_T}{P_S}$ is consequently
\begin{equation}
r=16 \epsilon_{CMB}.
\end{equation}

The scalar spectral index is given in terms of the slow-roll parameters
\begin{equation}
n_S=1-6 \epsilon_{CMB}+2 \eta_{CMB}.
\end{equation}
A scale invariant scalar spectrum corresponds to $n_S=1$.
For the cases that we will be interested in, $\epsilon_{CMB}\ll \eta_{CMB}$, so
$n_S\simeq 1+2 \eta_{CMB}$, and $\eta(\phi)\simeq m_p^2
\left(\frac{V''}{V}\right)$.

The running parameter of the scalar spectral index $\alpha=\frac{d n_S}{d \ln k}$ is given by
\begin{equation}
\alpha= -16\epsilon \eta+24 \epsilon^2+2\xi^2.
\end{equation}
In most of the cases that we will be interested in, both $\epsilon$ and $\eta$ are small so
$\alpha\simeq 2\xi^2$.

The tensor spectral index is $n_T\simeq -2 \epsilon_{CMB}$, which for the cases that we are interested in is
quite small. For the tensor perturbations a vanishing spectral index means scale invariance,
hence, when the slow-roll parameter $\epsilon$ is small the tensor index is scale invariant to a very good accuracy.

Generalizing to a multi field scenario, \cite{lythriotto,racetrack} the slow-roll parameters are defined as:
\begin{eqnarray}
\epsilon &=& \frac{1}{2}\left(\frac{g^{ab}\partial_a V \partial_b V}{V^2}\right)\\
\eta &=& \min \left\{\text{Eigenvalues}
\left(\frac{g^{ab}\partial_c \partial_b V-\Gamma^{\ \ d}_{cb}\partial_d V}{V}\right)\right\}
\end{eqnarray}
Here $g_{ab}$ is the field space metric  and $\Gamma^{\ \ c}_{ab}$ is the corresponding Christoffel symbol. Obviously, at an extremum the terms containing the Christoffel symbols vanish.

\subsection{Topological (hilltop) inflation}

We will be interested in a realization of inflation where a field is  ``trapped" near a local maximum of its potential, the so called ``topological inflation" \cite{guendelman,linde,vilenkin}.  A field configuration interpolating between two minima extends over a local maximum of the potential. The field configuration can, but does not necessarily have to, be generated by some symmetry and then it may have a topological origin as in the case of  monopoles or domain walls. For example, a domain wall solution that is protected by symmetry can form and under certain conditions its core will inflate eternally, thus creating our observed homogeneous and isotropic universe. Inflation ends in a certain patch of the universe when the scalar field falls from the top of the potential towards one of the minima.

The constraints that allow a solution with an inflating core can be recast as conditions on the potential. The width $|\phi_2-\phi_1|$ of the feature has to be larger than a a certain minimal width and the curvature of the potential at the maximum has to be small enough.
More concretely, let us denote the values of the fields at the minima by $\phi_1,\phi_2$ and the value of the field at the maximum by $\phi_{max}$. We can express the conditions for inflation as
\begin{eqnarray}
V(\phi_{max})&>&0 \\
\phi_2-\phi_1 &\gtrsim& m_{p} \\
0> \eta &\geq& -\mathcal{O}(10^{-2})
\end{eqnarray}
The condition on $\eta$  is needed to guarantee at least $60$ e-folds of inflation and to get rough agreement with CMB observations. The spectral index for the cases we consider is  $n_s=1+2\eta_{CMB}-6\epsilon_{CMB} \simeq 1+2\eta_{CMB}$. The CMB observations currently  prefer $n_s \simeq 0.95$ at about $60$ e-folds before the end of inflation. One expects that $|\eta_{CMB}|$ is somewhat different than $|\eta|$ at the extremum, however, generically,  for $|\eta|\gtrsim \mathcal{O}(10^{-1})$  the spectral index is in conflict with the data. Inflation models of this class have a small slow-roll parameter $\epsilon(\phi_{max})$ and
in them inflation is eternal.

This last property  relaxes considerably the problem of setting up appropriate initial conditions for inflation. In this case, it is enough that there will be some small and finite probability to enter into the eternal inflation regime.

In SUGRA models the chiral superfield space is a  K\"ahler space \cite{wessbagger}.
The metric in field space and the christoffel symbols can be expressed in terms of the real K\"ahler potential $K$:
\bea
g_{i \overline{j}}=g_{\overline{ i} j}=K_{i \overline{ j}} = \partial_i \partial_{\overline j}K \\
\Gamma^{\ \ k}_{ij}=K^{k \overline n}\partial_j \partial_i \partial_{\overline n} K.
\eea
The mixed holomorphic and anti-holomorphic Christoffel symbols vanish.
The slow-roll parameters have the following form
\bea
\epsilon &=&\frac{K^{i \overline j}\partial_i V \partial_{\overline j} V}{V^2} \\
\eta &=&\min \left\{\text{Eigenvalues}
\left(\begin{array}{cc} \vspace{.05in}
K^{i \overline m} N_{\overline m j} & K^{{i} \overline m} N_{\overline{m} \overline{j}} \\
K^{\overline{i}  m} N_{m  j } & K^{\overline i  m} N_{  m  \overline j}
\end{array}\right)\right\}
\eea
with
\bea
N_{i\overline j}&=&\frac{\partial_i \partial_{\overline j} V}{V}\cr
N_{i  j}&=& \frac{\partial_i\partial_{ j}V-\Gamma_{i j }^{\ \  k}\partial_{{k}}V }{V}.
\eea
In a multi-field space the extremum is not necessarily a maximum, rather it can be a saddle point. At the extremum $\epsilon=0$, however, since in this case there are several directions, $\eta$ is determined by the steepest direction and the constraint is still $0>\eta>-\mathcal{O}(10^{-2})$.

\subsection{Generic difficulties in designing small field SUGRA models of modular inflation}

As we have already explained one key reason favoring small field models is that the potentials for moduli fields are generically  exponentials that decay towards the outer region of moduli space and vanish in the weak coupling, decompactification limit. In the generic case all the slow-roll parameters at a generic point in moduli space will be of order one and inflation will be blocked. At specific points it is possible to avoid this conclusion. If one considers a small finite sum of exponentials then it is possible that the potential can be approximated as a polynomial in some region. One may hope that by fine tuning the parameters it might be possible to get  a large enough and flat enough region which can support inflation. Obviously the polynomial approximation (Taylor expansion) is valid in a limited range. Another logically possible way to design models is to consider a very large number of terms in the potential for a few fields or a very large number of fields. Then it seems likely that with enough tuning a potential that can support inflation can be designed. We will not pursue this approach in this paper.

We wish to illustrate here that designing successful models of modular inflation is much harder than it seems. In the next section we find the underlying reason and prove that with the perturbative K\"ahler potentials that we consider in this section such a construction is mathematically impossible. However, it is very useful to see in detail how the simple attempts fail. This will also explain why models that use the uplift term overcome many of the difficulties that we point out.

Let us briefly recall some properties of the moduli potentials in
the perturbative outer region of moduli space of string theory, where
the string coupling is weak and the values of the other moduli
(for example the volume of the compact space) are large. Realistic
models, for example in flux compactifications of string theory, usually
include an effective $N=1$ supergravity (SUGRA) theory below the
string scale $M_{s}=10^{-2}m_{p}$ and supersymmetry (SUSY) breaking
in some hidden sector at an intermediate scale $M_{I}=10^{-7} m_p$. The
most extensively studied class of models are those based on type IIB
string theory and follow the original discussion of \cite{GKP,KKLT} as reviewed in \cite{grana,douglaskachru}.
One expects a similar discussion to hold in other
string theories but since these have not been developed to the same
extent we will focus for the purpose of the current discussion on the type IIB case.

After compactification on a Calabi-Yau orientifold and for energies
well below the string scale and the compactification scale (i.e. the
Kaluza-Klein scale) the theory should take the form of an $N=1$ supergravity.
Thus the potential for chiral scalars~\footnote{For our purposes these would be the complete set of moduli fields
of the theory which describe the size and shape of the compact manifold
and the dilaton which sets the size of the string coupling.} would have to take the standard form,
\begin{equation}
V_{SUGRA}=e^{K}\left(K^{i\overline{j}}D_{i}WD_{\bar{j}}\overline{W}-3|W|^{2}\right),\label{vsugra}
\end{equation}
Here $K$, the K\"ahler potential, is a real analytic function of the
moduli and $W$, the superpotential,  is a holomorphic function of the
moduli. The K\"ahler derivative $D_{i}W\equiv\partial_{i}W+K_{i}W$
with $K_{i}=\partial_{i}W$ etc. $K_{i\bar{j}}=\partial_{i}\partial_{\bar{j}}K$
is the K\"ahler metric.

Let us assume for simplicity that the compactification manifold has
only one K\"ahler modulus which determines its overall volume. In
order to have a realistic theory wherein the cosmological constant
is tunable to its observed value, there should be on the order of
$10^{2}$ complex structure moduli. The K\"ahler potential is \begin{equation}
K=-3\ln(T+\bar{T})-\ln(S+\bar{S})-\ln k(z,\bar{z})\label{K}\end{equation}
where $T$ is the volume modulus $S$ is the dilaton and $k$ is
a real analytic function of the complex structure moduli $z$. The superpotential
is
\begin{equation}
W=A+BS+\sum_{i}A_{i}e^{-a_{i}T}.
\label{sumsp}
\end{equation}
where $A,B$ are holomorphic functions of the complex structure moduli.
The first two terms in this superpotential come from the internal
fluxes \cite{GKP} and the last set of terms in \cite{KKLT} (KKLT) arise from NP
effects (either from string instantons or gaugino condensation). Now
the original procedure of KKLT was to first ignore the  NP  terms in which case the global minimum of the potential is at
$F_{S}=F_{z}=0$. Since $W$ in this case is independent of $T$ there
is no potential for $T$. KKLT then fix $S,z$ at the points which
solve the F-term conditions of the previous sentence. While for the
purpose of just demonstrating the existence of a minimum for $T$
this strategy is reasonable, as a method of actually evaluating the
correct effective potential for $T$ it is incomplete. The reason
is that the actual superpotential now does have $T$-dependent terms
and although in the large volume regime the NP terms
are small, the entire potential is proportional to these terms and
would vanish in their absence. So the two stage procedure which solves
the above F-term conditions in the absence of the NP terms and then
derives the potential for $T$ by using
\begin{equation}
K=-3\ln(T+\bar{T}),\, W=W_{0}+\sum_{i}A_{i}e^{-a_{i}T},
\label{eq:toy}
\end{equation}
misses terms in the potential which are of the same order as the
ones which are being kept (for a detailed discussion see \cite{sda1}).
As shown in \cite{bda2} such a potential will not
have a SUSY breaking minimum with a zero or positive CC. What KKLT
did was to add a term to the potential (a so-called uplift term) which
may arise if the original string theory (ten-dimensional) background
had Dbar branes. However, from a four dimensional stand point this
would represent an explicit breaking of SUSY and the quantum
effective potential is liable to become quartically sensitive to the
ultraviolet physics. Nevertheless all (outer region) closed string
moduli inflation models have depended on this uplift term (for example, in \cite{racetrack,cq}). As will
become clear from the discussion of section~\ref{topinflation} (see also \cite{Badziak:2008yg})
these models just test this uplift term and do not appear to have
much to do with the actual supergravity part of the potential~\footnote{There is also the possibility of adding D-terms to the potential.
However this involves using open string moduli and we will defer a
discussion of such theories to future work.
}.

The moduli potential (coming from (\ref{K})(\ref{sumsp})) is actually
much more complicated than what is obtained from the toy model (\ref{eq:toy}).
In fact as shown in \cite{sda1} the theory can actually have
metastable minima with positive or zero cosmological constant. So
one might ask whether the actual supergravity potential for closed
string moduli coming from string theory can admit a realistic inflationary
cosmology. We will address this question in the next section.

Let us begin to explain why simple minded attempts to design small field inflationary models fail. For the purpose of the simple introductory discussion we will take a sum as in  eq.(\ref{eq:toy}) that contains only a few terms and we will take the coefficients $A_i$ and $a_i$  to be real. This assumption does not restrict the models in an
essential way and does allow us to understand more clearly the main properties of the
potentials.  We will be interested in general in
the situation that $a_i (T+\overline{T}) \gg 1$, that is in the outer region of moduli space.
The exact form of the K\"ahler potential will not be very important for the
discussion. What is important is that  derivatives of the K\"ahler potential scale such that they are smaller than derivatives of the superpotential, for example,
$K_{TT} W < K_T W_T $. The potential for $T$ simplifies considerably under these assumptions.

In general
the potential is then given by
\begin{equation}
V=e^K \left[K^{T \overline{T}}\left(W_T \overline{W}_{\overline{T}}+ W_T K_{\overline{T}}
\overline{W} +W K_T \overline{W}_{\overline{T}}+  K_T K_{\overline{T}} W \overline{W}\right)-3 W \overline{W} \right].
\end{equation}
However, under the simplifying assumptions we can use the simple form  $V= e^K K^{T \overline{T}}W_T \overline{W}_{\overline{T}}$.
For example, we can approximate
\begin{equation}
V_T=e^K K^{T \overline{T}}\left[W_{TT} \overline{W}_{\overline{T}}\right],
\end{equation}
and $D_TW=W_T$.
This conclusion is not strictly valid when the constant term in the superpotential is
substantial.

The approximations are very helpful in considering properties of extrema for the potential (without
a constant term). A minimum is a point where  $W_T=0$, since at that point $F=0$, so SUSY is unbroken, $V_T=0$, and the extremum is a minimum.
At a  maximum $W_T\ne 0$ and $W_{TT}=0$. Similarly, if we wish to compute the ratio  $V_{TT}/V$, which is relevant to determination of the $\eta$ parameter in models of inflation, we could estimate it as
$V_{TT}/V\simeq W_{TTT}/W_T$. In practice, the approximations work very well and lead to a
substantial simplification in understanding the properties of inflationary potentials.

\subsubsection{Difficulty in designing models with large distances between extrema}

In general, the potential will posses minima and maxima. The
superpotential and all its derivatives vanish at infinity (if there is a true
constant then all the derivatives vanish). We will be interested in the distance between some inner minimum and the adjacent
maximum as an estimate of the width of
the region near the maximum.

Assume that
\begin{align}
W_{TT}(T_{\rm max})& =0 \\
W_{T}(T_{\rm min}) & =0,
\end{align}
$|a T_{\rm max}|\gg 1$, $|a T_{\rm min}|\gg 1$ where $a$ is a representative exponent
in $W$. Define $\Delta T=T_{\rm max}-T_{\rm min}>0$. Now, let us show that $|a_i \Delta T|$ can be at most order unity, and that we cannot
have $a_i \Delta T \gg 1$. The idea is the following: all the exponentials that
participate in creating the minimum are of the same order $a$, and have similar
exponents and prefactors. Over a range $|a \Delta T| \gg 1$, only one exponent will
survive, and it cannot provide a maximum.
\begin{align}
W_{TT}(T_{\rm max}) & = \sum a_i e^{-a_i\Delta T} a_i A_i e^{-a_iT_{\rm min}}=0 \\
W_{T}(T_{\rm min}) & = \sum a_i A_i e^{-a_iT_{\rm min}}=0
\end{align}
Denote $x_i = a_i A_i e^{-a_iT_{\rm min}}$ and recall that we can assume that all terms are of order one
without loss of generality, and denote $c_i=a_i e^{-a_i\Delta T}$
\begin{align}
 \sum c_i x_i=0 \\
\sum  x_i=0
\end{align}
If $|a \Delta T| \gg 1$ then the $c_i$ corresponding to the smallest $a_i\Delta T$ will be much larger than the others. In this case the corresponding $x_i$ has to vanish, in contradiction to the assumption that all the $x_i$ are of order one.

\subsubsection{Difficulty in designing models with a small value of $\eta$}

The maximum that we are looking for to allow the field to inflate off of it has a
(nearby) minimum accompanying it. We may expand
\begin{equation}
W_T(T_{\rm max})=W_T(T_{\rm min})+ \Delta T W_{TT}(T_{\rm min})+\cdots
\end{equation}
As we have showed in the previous subsection, the higher terms in the expansion are small and can be neglected.
\begin{equation}
W_{TT}(T_{\rm min})=W_{TT}(T_{\rm max})- \Delta T W_{TTT}(T_{\rm max})+\cdots
\end{equation}
Since $W_T(T_{\rm min})=0$ and $W_{TT}(T_{\rm max})=0$,
\begin{equation}
W_T(T_{\rm max})=\Delta T W_{TT}(T_{\rm min})=-(\Delta T)^2 W_{TTT}(T_{\rm max})
\end{equation}
and hence
\begin{equation}
\frac{W_{TTT}(T_{\rm max})}{W_T(T_{\rm max})}=- \frac{1}{(\Delta T)^2}
\end{equation}

For perturbative K\"ahler potentials we expect $K^{T\overline{T}} \sim (T+\overline T)^2$ so the value of $\eta$ scales approximately as
\begin{equation}
\eta_{\rm can}\sim (T_{\rm max})^2 \frac{W_{TTT}(T_{\rm max})}{W_T(T_{\rm max})}=
-\frac{(T_{\rm max})^2}{(\Delta T)^2}.
\end{equation}
If $T_{\max}> \Delta T$ as expected in outer region models, the value of $\eta$ cannot be small.

\section{Topological (hilltop) SUGRA inflation}
\label{topinflation}

\subsection{A No-go theorem for a logarithmic K\"ahler potential}

We will now prove for SUGRA models with a single complex field $T$ that if $K=-A\ln \tbt$ with $A\leq 3$, and the superpotential is an arbitrary regular holomorphic function then it is not possible to find an extremum that satisfies the following conditions,
\begin{eqnarray}
\label{posv}
&1.& \ V(T_0,\overline T_0)>0  \\
\label{etacond}
&2.& \ -\frac{1}{100}\lesssim \eta <0.
\end{eqnarray}

We prove our result by showing that the trace of the two dimensional $\eta$ matrix is  negative and that its magnitude is at least of order unity. By showing this we would have proved that the most negative eigenvalue of the $\eta$ matrix is too large and cannot satisfy condition (\ref{etacond}). This was also proved independently by \cite{Badziak:2008yg}.

We begin by listing the potential and its derivatives for the single field case,
\begin{eqnarray}
V & = & e^{K}\left(|D_{T}W|^{2}K^{T\overline{T}}-3|W|^{2}\right)\label{eq:V}\\
\partial_{T}V & = & e^{K}\left(D_{T}^{2}WD_{\overline{T}}\overline{W}K^{T\overline{T}}-2D_{T}W\overline{W}\right)
\label{eq:dV}\\
\nabla_{\overline{T}}\partial_{T}V & = & e^{K}\left(-R_{T\overline{T}T\overline{T}}(K^{T\overline{T}})^2~D_{T}W D_{ \overline T} \overline W
+ D_{T}^{2}W D_{ \overline T}^2 \overline W K^{T\overline{T}}-2K_{T\overline{T}}|W|^{2}\right).
\label{eq:d2V}
\end{eqnarray}
The K\"ahler derivatives are covariant so when acting on tensors they include contributions from the Christoffel symbols. At an extremum $\nabla_{\overline{T}}\partial_{T}V=\partial_{\overline{T}}\partial_{T}V$. The Riemann  tensor  is $R_{T\overline{T}T{\overline{T}}}=K_{T\overline{T}{T\overline{T}}}-K^{T\overline{T}} K_{T\overline{T}T} K_{\overline{T}T\overline{T}}$.
At the extremum $T=T_0$  $V(T_0)>0$ so SUSY is broken and $D_T W \ne 0$. In this case  the extrmality condition from eq.~(\ref{eq:dV}) is solved by
\begin{equation}
\label{solA}
|D_{T}^{2}W|K^{T\overline{T}}=2|W|.
\end{equation}
Substituting eq.~(\ref{solA}) into eq.~(\ref{eq:d2V}) gives
\begin{equation}
\partial_{\overline{T}}\partial_{T}V|_{T_0}= e^{K(T_0,\overline{T}_0)}\left(-R_{T\overline{T}T\overline{T}}(K^{T\overline{T}})^2~D_{T}W
D_{ \overline T} \overline W + 2K_{T\overline{T}}|W|^{2}\right)|_{T_0}.
\label{eq:d2V2}
\end{equation}

The superpotential is a regular  holomorphic function and so can be expanded $W=\sum b_{i}(T-T_{0})^{i}$ (the $b_i$ are complex). Then,
\begin{eqnarray}
W({T_0})&=&b_{0} \nonumber \\
D_{T}W|_{T_0}&=&b_{1}-\frac{Ab_{0}}{T_{0}+\overline{T}_{0}} \nonumber \\
K_{T\overline{T}}|_{T_0}&=& \frac{A}{(T_0+\overline{T}_0)^{2}} \nonumber \\
(K^{T\overline{T}})^2 R_{T\overline{T}T\overline{T}}&=& \frac{2}{A}.
\end{eqnarray}
We may define $\left|B\right|^{2}=D_{T}W
D_{ \overline T} \overline W K^{T\overline{T}}|_{T_0}$, \[\left|B\right|^{2}=\left|b_{1}-\frac{Ab_{0}}{T_{0}+\overline{T}_{0}}\right|^2 \frac{\left(T_{0}+\overline{T_{0}}\right)^{2}}{A}.\] In terms of $|B|^2$,
\begin{eqnarray}
\label{vt}
V(T_0,\overline T_0) & = & e^{K(T_0,\overline{T}_{0})}\left(|B|^{2}-3|b_{0}|^{2}\right)\\
\label{vttbar}
\partial_{\overline{T}}\partial_{T}V|_{T_0} & = & -2e^{K(T_0,\overline{T}_0)}\left(|B|^{2}-|b_{0}|^{2}A\right)\frac{1}{\left(T_{0}+\overline{T_{0}}\right)^{2}}.
\end{eqnarray}
From condition (\ref{posv}) it follows that
\begin{equation}
\label{posVexp}
|B|^{2}-3|b_{0}|^{2}>0.
\end{equation}

The $\eta$ matrix for our single (complex) field case is
\begin{equation}
\label{etaT}
\eta= \left(\begin{array}{cc}\frac{K^{\overline T T} V_{T\overline{T}}}{V} &  \frac{K^{\overline T T} V_{{T}{T}}}{V}\\  \frac{K^{\overline T T} V_{\overline{T}\overline{T}}}{V} & \frac{K^{\overline T T}  V_{\overline{T}{T}}}{V}\end{array}
\right).
\end{equation}
We substitute eq.~(\ref{vttbar}) into eq.~(\ref{etaT}) and evaluate the trace
\begin{eqnarray}
\label{traceeta}
\text{Tr}\ \eta &=& 2\frac{(T_0+\overline T_0)^2}{A }\frac{V_{T \overline T}}{V}|_{T_0} \nonumber \\
&=&-\frac{4}{A}\frac{|B|^{2}-A|b_{0}|^{2}}{|B|^{2}-3|b_{0}|^{2}} \nonumber \\
&=& -\frac{4}{A}\left(1+ (3-A) \frac{|b_{0}|^{2}}{|B|^{2}-3|b_{0}|^{2}}\right).
\end{eqnarray}
Combining eq.~(\ref{posVexp}) with the fact that $A\leqslant 3$ we find that  $$(3-A) \frac{|b_{0}|^{2}}{|B|^{2}-3|b_{0}|^{2}}\geqslant 0. $$
Then from eq.~(\ref{traceeta}) we find that
\be
\text{Tr}\ \eta \leqslant -\frac{4}{A} \hspace{.2in} \text{for}\ \ 0<A\leqslant  3.
\ee
Since $-\frac{4}{A} \leqslant -4/3\  \text{for}\ \ 0<A\leqslant  3$ it is clear that the most negative eigenvalue of the matrix is nowhere near the desired small value $ \gtrsim - 1/100 $.

\subsection{Expansion of the K\"ahler potential}
In the previous section we have derived a necessary condition for viable inflation in a supergravity theory of a single chiral field $T$.
We have found that
\begin{equation}
\label{trcond}
Tr[\eta]=2\frac{K^{T\overline{T}}\nabla_{\overline{T}}\partial_TV}{V}\bigl|_0=2\frac{e^K}{V}\left[-R_{T\overline{T} T \overline{T}}(K^{T\overline{T}})^3+2|W|^2\right]{\biggl|_0} >-\mathcal{O}(10^{-2}).
\end{equation}
If the inequality does not hold at the extremum, at least one direction in field space is too steep and slow-roll inflation does not occur.
Since $|W|^2, K^{T\overline{T}}$ are positive semi-definite whether condition~(\ref{trcond}) is satisfied or not depends on the Riemann curvature tensor. If
\begin{equation}
\label{trcond2}
2|W|^2\gtrsim R_{T\overline{T} T \overline{T}}(K^{T\overline{T}})^3
\end{equation}
then it is no longer true that at least one eigenvalue has to be large and negative.
Obviously, if
 \begin{equation}
R_{T\overline{T} T\overline{T}}\leq 0.
 \end{equation}
then condition~(\ref{trcond2}) is automatically satisfied. We conclude that if the field space is flat, as in the case that K\"ahler potential is canonical, or if it is negatively curved, then condition~(\ref{trcond}) is satisfied.

The expression for the Riemann tensor is:
 \begin{eqnarray}
R_{T\overline{T}T{\overline{T}}}=K_{T\overline{T}{T\overline{T}}}-K^{T\overline{T}} K_{T\overline{T}T} K_{\overline{T}T\overline{T}}.
 \end{eqnarray}
Since the second term is negative semi-definite from the reality of $K$, a small or negative fourth derivative of the K\"ahler potential
 \begin{equation}
\label{4kder}
 K_{T \overline{T} T \overline{T}} \leq 0,
 \end{equation}
is enough to satisfy condition~(\ref{trcond}).

Expanding a general $K$ about an extremum at $T_0$ and redefining $T$ to be the deviation from $T_0$, only terms up to fourth order are relevant:
 \begin{eqnarray}
K(T,\bar T)=a_0+a_{10} T+a_{01} \bar T+a_{20} T^2+a_{11}T \bar T+a_{02}\bar T^2+a_{30} T^3+a_{21}T^2 \bar T\cr
+a_{12} T \bar T^2+a_{03} \bar T^3+a_{40} T^4+a_{31} T^3 \bar T+a_{22}T^2 \bar T^2+a_{13} T \bar T^3+a_{04} \bar T^4
\end{eqnarray}
The reality of $K$  implies that $a_{ij}=a^*_{ji}$. Using the K\"ahler transformation
$
K(T, \bar T)\rightarrow K(T, \bar T)+f(T)+\bar f(\bar T)$,
$ W(T)\rightarrow W(T)e^{-f(T)}$ allows us to set $a_{0i}=0$, so
\begin{eqnarray}
K(T,\bar T)&=&a_{11}T \bar T+a_{21}T^2 \bar T+a_{12} T \bar T^2+a_{31} T^3 \bar T+a_{22}T^2 \bar T^2+a_{13} T \bar T^3,
\end{eqnarray}
while the superpotential can still be expanded in a Taylor series with arbitrary coefficients $b_i$, $W=\sum_i b_i T^i$.
In terms of the expansion parameters condition~(\ref{trcond2}) now reads
\begin{equation}
|b_0|^2\geq \frac{2}{a_{11}^4}\left(a_{11}a_{22}-|a_{12}|^2\right).
\end{equation}
It is now clear that condition~(\ref{4kder})  simply corresponds to $a_{22}\leq 0$. In particular if $a_{22}=0$ it is automatically satisfied.

We conclude that the K\"ahler potential has to be significantly different from its form in perturbative string theory to avoid the results of the previous subsection.

\subsection{Justifying the string theoretic origin of the single field SUGRA models}
The arguments of the previous subsections  apply to SUGRA potentials with one complex
field $T$. In string theory, however, there are many complex moduli
fields which are coupled together in a complicated way. We will argue
below that when the K\"ahler potential (and
hence the metric) and the superpotential for the moduli are a direct sum of the one for the
would be inflaton modulus (which we call $T$) and the other moduli - then the
same conclusion holds in the perturbative region where the moduli
take large values. We believe that similar conclusion holds in the outer region of moduli space also for more general multifield theories. Our conclusion changes for smaller values of the moduli, i.e. when we are in the central region of moduli space.

The general expressions for the potential and its first two derivatives
are as follows.
\begin{eqnarray}
V & = & e^{K}(D_{i}WD_{\bar{j}}\overline{W}K^{i\bar{j}}-3|W|^{2})\label{eq:P}\\
\partial_{k}V & = & e^{K}(D_{k}D_{i}WD_{\bar{j}}\overline{W}K^{i\bar{j}}-2D_{k}W\overline{W})\label{eq:dP}\\
\nabla_{l}\partial_{k}V & = & e^{K}(D_{l}D_{k}D_{i}WD_{\bar{j}}\overline{W}K^{i\bar{j}}-D_{l}D_{k}W\overline{W})\label{eq:ddP}\\
\nabla_{\bar{l}}\partial_{k}V & = &
e^{K}(-R_{k\bar{l}i\bar{m}}D_{n}WD_{\bar{j}}\overline{W}K^{i\bar{j}}K^{n\bar{m}}+K_{k\bar{l}}D_{i}WD_{\bar{j}}\overline{W}K^{i\bar{j}}-D_{k}WD_{\bar{l}}\overline{W}\nonumber
\\
 &  & +D_{k}D_{i}WD_{\bar{l}}D_{\bar{j}}\overline{W}K^{i\bar{j}}-2K_{k\bar{l}}W\overline{W}).
 \label{eq:ddbarP}
 \end{eqnarray}
In the above $\partial_{i}$ denotes differentiation with respect
to a chiral scalar $\phi^{i}$, $K_{i}=\partial_{i}K$ etc. and
\begin{eqnarray}
D_{i}X_{j}&=&\nabla_{i}X_{j}+K_{i}X_{j} \cr
\nabla_{i}X_{j}&=&\partial_{i}X_{j}-\Gamma_{ij}^{k}X_{k} \cr
\Gamma_{ij}^{k}&=&K^{k\bar{l}}\partial_{i}K_{j\bar{l}} \\
R_{i\bar{j}k\bar{l}}&=&K_{m\bar{l}}\partial_{\bar{j}}\Gamma_{ik}^{m}. \nonumber
\label{eq:Kahlergeom}
\end{eqnarray}

As before we find the conditions for inflation by evaluating
the inflationary parameters near an extremum which is in general a
saddle point. At the extremum $\partial_{k}V=0$ and we get from (\ref{eq:dP})
\begin{equation}
D_{k}^{}D_{i}WD_{\bar{j}}\overline{W}K^{i\bar{j}}|_{0}=2D_{k}W\overline{W}|_{0}
\label{eq:extr}
\end{equation}
Now since the potential is positive at this extremum (in order to
have inflation) supersymmetry is necessarily broken. Let us call this
broken direction $T$. i.e. we assume that
\begin{equation}
 D_{T}W\ne0,\,  D_{i}W=0,\, i\ne T.\label{eq:Fterms}\end{equation}

Let us also assume a direct product space so that the K\"ahler potential
takes the form,\begin{equation}
K=K_1(T,\overline{T})+K_2(\phi^i,\bar{\phi^i})\label{eq:Kdirect}\end{equation}
with $i\ne T$. Then
the corresponding metric is block diagonal with \begin{equation}
K_{i\overline{T}}=K_{T\bar{i}}=0.\label{eq:nondiag}\end{equation}
This would be the case if $T$ is the dilaton of string theory or the volume modulus in a compactification of string theory
on a manifold with one K\"ahler modulus. It would be true also for
the K\"ahler and complex structure modulus in orbifold compactifications.
Now choose $k\ne T$ in eq.~(\ref{eq:extr}) and use eq.~(\ref{eq:nondiag})  to get
\begin{equation}
D_{k}D_{i}WD_{\bar{j}}\overline{W}K^{i\bar{j}}=\sum_{i,j\ne T}D_{k}D_{i}WD_{\bar{j}}\overline{W}K^{i\bar{j}}+D_{k}D_{T}WD_{\overline{T}}\overline{W}K^{T\overline{T}}=0
\end{equation}
at the extremum.
Using eq.~(\ref{eq:Fterms}) this
gives
\begin{equation}
D_{T}D_{k}W|_0=D_{k}D_{T}W|_{0}=0.
\label{eq:dkTW}
\end{equation}
Now choose $k=T$ in (\ref{eq:extr}). Using again eqs.~(\ref{eq:Fterms}), (\ref{eq:nondiag})
we have
\begin{equation}
|D_{T}^{2}W|_{0}=2K_{T\overline{T}}|W|_{0}\label{eq:dtsquared}.
\end{equation}
We further observe that in eq.~(\ref{eq:ddbarP}) (with $k,l=T$) $R_{T\overline{T}i\bar{m}}K^{i\bar{j}}K^{n\bar{m}}D_{n}WD_{\bar{j}}\overline{W}|_{0}=
R_{T\overline{T}T\overline{T}}(K^{T\overline{T}})^{2}|D_{T}W|_{0}^{2}$
and that the fourth term inside the parenthesis vanishes because of
eq.~(\ref{eq:dkTW}) so that we have
\begin{equation}
\nabla_{\overline{T}}\partial_{T}V=e^{K}\{-R_{T\overline{T}T\overline{T}}(K^{T\overline{T}})^{2}|D_TW|^2+2K_{T\overline{T}}|W|^{2}\},\label{eq:PTTbar}
\end{equation}
which is just the expression we had previously for the single field case. However,
to draw the same conclusions we have to argue that the matrix of
second derivatives is block diagonal in the $T\bar{,T}$ sector.

First consider the expression for $\nabla_{\bar{l}}\partial_{T}V|_{0}$
with $l\ne T$ (see eq.~(\ref{eq:ddbarP})).
\begin{eqnarray}
\nabla_{\bar{l}}\partial_{T}V & = &
e^{K}(-R_{T\bar{l}i\bar{m}}D_{n}WD_{\bar{j}}\overline{W}K^{i\bar{j}}K^{n\bar{m}}
+K_{T\bar{l}}D_{i}WD_{\bar{j}}\overline{W}K^{i\bar{j}}-D_{T}WD_{\bar{l}}\overline{W}\nonumber\\
 &  & +D_{T}D_{i}WD_{\bar{l}}D_{\bar{j}}\overline{W}K^{i\bar{j}}-2K_{T\bar{l}}W\overline{W}).
 \end{eqnarray}
All terms except the first term inside the parenthesis vanish because of eqs.~(\ref{eq:Fterms}), (\ref{eq:nondiag})
and (\ref{eq:dkTW}). The only potentially non-vanishing term comes
from the first term,
\begin{equation}
\nabla_{\bar{l}}\partial_{T}V  =
-e^{K}R_{T\bar{l}i\bar{m}}D_{n}WD_{\bar{j}}\overline{W}K^{i\bar{j}}K^{n\bar{m}}
=
-e^{K}R_{T\bar{l}T\overline{T}}D_{T}WD_{\overline{T}}\overline{W}K^{T\overline{T}}K^{T\bar{T}},
\end{equation}
where the second equality is again due to eq.~(\ref{eq:Fterms}) and eq.~(\ref{eq:nondiag}).
$R_{T\bar{l}T\overline{T}}$ vanishes
because the direct product nature of the space (\ref{eq:Kdirect}), so we have:
\begin{equation}
\nabla_{\bar{l}}\partial_T V=\nabla_{\overline{T}}\partial_l V=\nabla_{T}\partial_{\bar{l}} V=\nabla_{l}\partial_{\overline{T}} V=0, \quad \text{for}\ l \neq T.
\end{equation}
Thus the submatrix of $ $holomorphic and anti-holomorphic derivatives
is block diagonal with a $T,\overline{T}$ block. Now let us look at the
holomorphic- holomorphic sector (and its complex conjugate) i.e. eq.~(\ref{eq:ddP}),
\begin{equation}
\nabla_{l}\partial_{k}V  = e^{K}(D_{l}D_{k}D_{i}WD_{\bar{j}}\overline{W}K^{i\bar{j}}-D_{l}D_{k}W\overline{W})
\end{equation}
The relevant mixed term (with say $l\neq T$
and $k= T$) has a potentially non vanishing term (under the conditions
(\ref{eq:Fterms}), (\ref{eq:nondiag})) of the form $D_{l}D_{T}D_{T}WD_{\overline{T}}\overline{W}K^{T\overline{T}}$.
The last two factors are certainly non-vanishing. Evaluating the first
factor (after using the fact that the mixed connection factors are zero) we find:
\begin{eqnarray}
D_lD_TD_TW|_0 &=&\partial_lD^2_TW|_0+K_lD^2_TW|_0 \cr
&=&\partial_lD^2_TW|_0+2K_lK_{T\overline{T}}W|_0,
\end{eqnarray}
where in the second equality we used eq.~(\ref{eq:dtsquared}).
Now if the superpotential is a sum of holomorphic functions:
\begin{equation}
W=W_1(T)+W_2(\phi^l),
\end{equation}
the mixed derivative vanishes $\partial_{l}\partial_{T}W=0$.
This form of $W$ occurs, for example, when $T$ is identified with the
volume modulus $\tau$ and $l$ is either the dilaton or a complex
structure modulus.
Thus, the mixed holomorphic-holomorphic derivative is:
\begin{equation}
\nabla_l\partial_TV\simeq K_lWD_TW.
\end{equation}
It contains an extra factor of $K_l$ multiplying terms comparable to those in eq.~(\ref{eq:PTTbar}), so for large
moduli $\phi^l$ this term is expected to be subdominant.

Thus at least in the case where $T$ is identified with the volume modulus in compactifications with
just one K\"ahler structure the mass matrix is approximately block diagonal
in the perturbative large moduli regime. If $T$ is to be identified
with the dilaton $S$ or a complex structure modulus one needs additional
restrictions on the moduli superpotential.

For KKLT type models the superpotential is of the form $W=A+BS+\sum_i C_ie^{-a_i\tau}$ where $S$
is the dilaton, $\tau$ is the volume modulus and $A$ and $B$ are functions of the complex structure modulus, the suppression of the non-diagonal terms in the mass matrix can be checked and calculated explicitly.
The term involving $\partial_{T}\partial_{l}W$ is zero if $T$ is identified with the
volume modulus $\tau$ and $l$ is either the dilaton or a complex
structure modulus.

To summarize, under the conditions:
\begin{eqnarray}
K&=&K_1(T,\overline{T})+K_2(\phi^l,\bar{\phi}^l)\\
W&=&W_1(T)+W_2(\phi^l)\\
K_l &\ll& 1
\end{eqnarray}
then the $T,\overline{T}$ sector mass matrix decouples
from the rest of the moduli i.e. the mass matrix becomes (approximately) block diagonal
and the previous analysis remains valid. In particular the sum of
the eigenvalues of the $\eta$ matrix in the $T$ directions are less
than $-4/3$ unless $ R_{T\overline{T}T\overline{T}}(K^{T\overline{T}})^{2}\le 4/3$
which is certainly not the case for typical string theory moduli in the perturbative region as
discussed before.

\subsection{Models with a canonical K\"ahler potential}

We now wish to show that SUGRA models with a canonical K\"ahler potential can be constructed such that they satisfy all the required constraints for a viable inflation model. For small field models we may expand the superpotential in a Taylor series. Generally speaking, integrating out the heavy moduli makes the superpotential non-chiral (non-holomorphic). However if the coupling between the fields is small enough, the holomorphicity of the superpotential is preserved and the single field construction using a real $K(T,\overline{T})$ and a chiral $W(T)$ is valid.

We assume that the potential has an extremum at $T=T_0$, and that the width of the feature we wish to construct is $\Delta T= T_2-T_1$. The width has to be larger than about unity in reduced Planck units. The conditions for a good inflationary model  are:
\begin{eqnarray}
\label{eq:conditions1}
V_T(\tn, \btn), V_{\overline{T}}(\tn, \btn) =0 \\
\label{eq:conditions2}
V(\tn,\btn)>0 \\
\label{eq:conditions3}
|\eta|< \mathcal{O}(10^{-2})
 \\
\label{eq:conditions4}
|\Delta T| \gtrsim m_p.
\eea

We may choose without loss of generality $T_0=0$ by redefining $T$ to be the deviation from $T_0$.  We may  further choose $V(\tn,\btn)=1$ due to the freedom of rescaling the superpotential $W \rightarrow cW$ which rescales the potential as $V\rightarrow c^2 V$.  The conditions fixing  $T_0$ ,$T_1$, $T_2$ and $\eta$ are not affected by this rescaling.

The conditions (\ref{eq:conditions1}), (\ref{eq:conditions2}), (\ref{eq:conditions3}) on the extremum and on $\eta$ are local conditions involving only a finite number of derivatives of the superpotential at $T_0$. These can be translated to algebraic equations for the expansion coefficients of $W$. However, condition (\ref{eq:conditions4}) on the width of the feature is not a local condition. To be able to get tractable equations we turn it into a local condition. We know that points where $D_T W=0$ are minima of the potential. So we impose two  such minima at $T_1$ and $T_2$ at a prescribed distance from the extremum at $T_0$. At points where the K\"ahler derivative of the superpotential vanishes the potential is negative.
As it will turn out,  designing a ``good" Minkowski or deSitter minimum requires the addition of just one condition on $W$. Such an example will be given at the end of the section.

The equations $D_T W_{|T_1,T_2} =0$ are algebraic equations in the coefficients and the distance between minima. In practice, for simplicity, we choose symmetric minima $T_{1,2}=\pm y$,
\begin{eqnarray}
\label{eq:pconditions1}
V_T(0) &=& 0 \\
\label{eq:pconditions2}
V(0)&=&1 \\
\label{eq:pconditions3}
|\eta|&<&\mathcal{O}(10^{-2}) \\
\label{eq:pconditions4}
D_T W(\pm y)&=&0,\ \  y \sim 1  
\end{eqnarray}
Conditions~(\ref{eq:pconditions1},\ref{eq:pconditions2},\ref{eq:pconditions3},\ref{eq:pconditions4}) are translated into algebraic relations that the expansion coefficients of the superpotential satisfy. For simplicity, we consider real coefficients in the expansion of the superpotential.

Since 5 equations need to be solved, $W$ has to have at least 5 free parameters to allow a solution. If some of the equations are linearly dependent, for example, as in the case of a periodic potential a lesser number of free parameters is needed. However,  we have found that to satisfy all the constraints at least 6 free parameters are required  because for 5 parameters one gets an implicit relation of the form $y\simeq|\eta|^{-1/2}$ which cannot be satisfied for $y$ of order unity.

For
\begin{eqnarray}
K&=&T \bt \cr
W&=&b_0+b_1 T+b_2 T^2+b_3 T^3+b_4 T^4+b_5 T^5+\cdots,
\end{eqnarray}
eq.~(\ref{eq:pconditions1}) and eq.~(\ref{eq:pconditions2}) become algebraic equations for the expansion coefficients
\begin{eqnarray}
\label{eq:sol1}
2 b_1(b_0-b_2)&=&0, \\
\label{eq:sol2}
-3b_0^2+b_1^2&=&1.
\end{eqnarray}
Equation (\ref{eq:sol1}) can be solved in two ways, either $b_1=0$ or $b_0=b_2$. If $b_1=0$ then eq.~(\ref{eq:sol2}) cannot be satisfied so $b_0=b_2$. Since we are looking for a symmetric potential about $T_0=0$, we choose $b_0,b_2,b_4=0$ so that eq.~(\ref{eq:pconditions4}) is an even equation. This is obviously compatible with eqs.~(\ref{eq:sol1},\ref{eq:sol2}).
Then from eq.~(\ref{eq:sol2}) we find $b_1=1$. We now continue to determine the other parameters.

The $\eta$ matrix in this case is given by
\begin{equation}
\eta=
\left(\begin{array}{c c}
 2 b_0^2   & 6 b_3 b_1 - 2b_0^2\\
6 b_3 b_1 - 2b_0 b_2  &   2 b_0^2
\end{array}\right)= \left(\begin{array}{c c}
  0  & 6 b_3 \\
 6 b_3 &  0
\end{array}\right),
\end{equation}
and the value of most negative eigenvalue is in this case
\begin{eqnarray}
\eta&=&-6 |b_3|.
\end{eqnarray}
We may choose $b_3$ to be negative and then $\eta=6 b_3$.
The remaining equation determines $b_5$ in terms of $y$ and $\eta$.
Substituting $T_1,T_2=\pm y$  into (\ref{eq:pconditions4}) we obtain
 \begin{eqnarray}
y(b_1y+b_3y^3+b_5y^5)+b_1+3b_3y^2+5b_5y^4&=&0 \\
-y(-b_1y-b_3y^3-b_5y^5)+b_1+3b_3y^2+5b_5y^4&=&0.
\end{eqnarray}
whose solution is
\begin{eqnarray}
b_5=-\frac{1+y^2+\frac{1}{6} \eta~ y^2(3+y^2)}{y^4(5+y^2)}\simeq -\frac{1+y^2}{y^4(5+y^2)} .
\end{eqnarray}

In summary, for the case of a canonical K\"ahler potential,
\begin{eqnarray}
\label{design1}
W&=& T+(\eta/6)~ T^3+b_5 T^5 \\
\label{design2}
b_5&=& -\frac{1+y^2+\frac{1}{6} \eta\ y^2(3+y^2)}{y^4(5+y^2)},
\end{eqnarray}
The potential has a saddle point at $T=0$, a local maximum in the real direction that has small negative curvature $\eta$ and two negative minima at $T = \pm y$.
Equations (\ref{design1}) and (\ref{design2}) provide a starting point for designing models of topological SUGRA inflation, which we do  next.

In Table~\ref{tb1} we give several examples and calculate their predictions for the scalar spectral index $n_S$, for the scalar to tensor ratio $r$ and for the spectral index running parameter $\alpha$.
The superpotential is given by $W = T+(\eta/6)~  T^3+b_5 T^5 $. In the table we list $b_5$, half the distance between the minima $y$ and the value of $\eta$ at the maximum,
\begin{table}
\begin{center}
\begin{tabular}{l|c|c|c|c|c|c|}
  \cline{2-7}
  \ & $b_5$ & $y$ &$\eta$& $n_S$ & $r$ &$\alpha $  \\ \hline
   \hspace{-.085in}\vline\ I& -0.33 & 1 & -0.01 & .97 & $9\times 10^{-7}$ & $- 6.3\times 10^{-4}$ \\  \hline
    \hspace{-.085in}\vline\ II& -3.8 & 0.5 & -0.02 & .96 & $6\times 10^{-8}$ & $-3.4\times 10^{-4}$ \\ \hline
     \hspace{-.085in}\vline\ III& -1.95 & 0.6 & -0.005 & .97 & $1\times 10^{-7}$ & $-7.6\times 10^{-4}$\\ \hline
      \hspace{-.085in}\vline\ IV& -0.33 & 1 & -0.04 & .92 & $2 \times 10^{-7}$ & $-4.0\times 10^{-5}$ \\ \hline
       \hspace{-.085in}\vline\ V& -0.044 & 1.85 & -0.02 & .97 & $1\times 10^{-4}$ & $4.3\times 10^{-4}$\\ \hline
\end{tabular}
\end{center}
\caption{Models of  inflation and their CMB parameters}
\label{tb1}
\end{table}
In order to find the spectral index $n_s\simeq 1+2\eta_{CMB}$, the scalar to tensor ratio $r=16\epsilon_{CMB}$ and the running parameter $\alpha$  we consider inflating solutions that evolve along the real $T$ direction and numerically integrate  the equations to get to $60$ e-folds before the end of inflation. The running parameter $\alpha$ is evaluated using the approximate relation \cite{lythriotto} $\alpha=-2 \frac{V_i}{V} \frac{d\eta}{d\phi_i}$.

One can easily come up with numerous other examples of this sort.
Several aspects of the results are generic to all examples. First, $n_S\lesssim 1$
and the same $n_S$ can be obtained for different $\eta$. This is due to the fact that $\eta$ changes from its value at the extremum to $\eta_{CMB}$. Second, the amplitude of GW in these models is extremely small, beyond hope of detection in the foreseeable future. Third, the value of $\alpha$ is extremely small, beyond hope of detection in the foreseeable future. Perhaps this is due to a specific choice of parameters and some other choice may lead to a larger amplitude? In the next section we demonstrate that in the simple single field case the smallness of $r$ and $\alpha$ are robust and one always gets $r<10^{-4}$ and $\alpha < 10^{-3}$.

\FIGURE[t]{
\scalebox{1.5}{\includegraphics{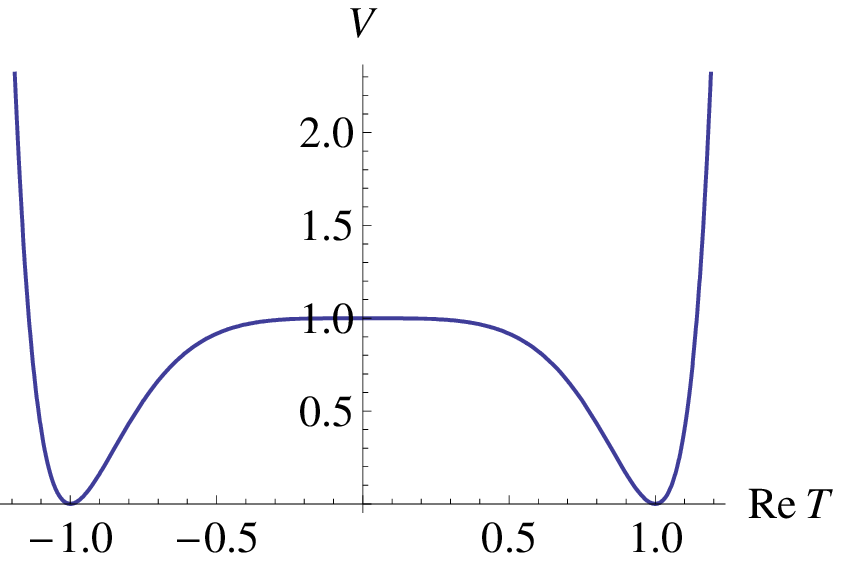}}
\caption{ \label{figure1} Shown is a graph of the potential resulting from example~\ref{goodmin} in the $Re T$ direction for $Im T=0$.}}

We now explain how to design a ``good" Minkowski or deSitter minimum in addition to the inflationary features. The change is minimal, instead of the condition $D_T W=0$ one requires $V_T(y)$, $V(y)=0$ and $D_TW(y)\neq 0$. This increases the number of conditions by one to six.  Adding just one more parameter to the superpotential $b_7$ is enough to allow design of such models. For simplicity we keep the potential symmetric around the origin so $b_{2n}=0$. To demonstrate the validity of the procedure we present a numerical example with $\eta=-0.01$ and $y=1$,
\begin{equation}
\label{goodmin}
b_1=1,\quad b_3=-0.01/6, \quad b_5=-0.158142319, \quad b_7=0.065535647,
\end{equation}
with the corresponding observables:
\begin{equation}
n_S=.97, \quad r=2.3 \times 10^{-6}, \quad \alpha=-6.1 \times 10^{-4}.
\end{equation}
The value of the potential at the minimum (in this example $V(T=1)\sim 10^{-9}$) is much more sensitive than other quantities to the values of the parameters. Tuning the parameters away by as little as one percent gives the same inflationary observables and a  deSitter minimum with a large value of $V(T=1)\sim 0.2$.

\section{Suppression of gravitational waves and spectral index running in hilltop modular inflation}
\label{GW}

In this section we determine the phenomenological consequences of models of modular inflation of the class that was described in the previous section and show that they predict a red spectrum of scalar perturbations and a characteristic suppression of the tensor perturbations and the running of the spectral index. Consequently, despite the high scale of inflation, a detection of primordial tensor anisotropies or spectral index running in CMB observations in the foreseeable future \cite{wmap,planck} will rule out this entire class of models .

We use here simplified bosonic models because they are easier to analyze and capture the essential features of the SUGRA models. Specifically, we consider models with a single real bosonic field where an analytic derivation of the results and their dependence on the various parameters is possible. These models will allow us to explain the reasons for the suppression of the tensor perturbations and running of the spectral index.

Let us consider the potential
\begin{equation}
V(\phi)= \Lambda^4 \left(1-a_2 \phi^2-a_3 \phi^3 -a_4 \phi^4 +\cdots\right).
\end{equation}
We assume that the potential has a maximum at
$\phi=0$, so $a_2$ has to be positive in our conventions, however, the signs of the other coefficients are not restricted. We have normalized the field $\phi$ in (reduced) Planck units. We assume that the field slowly rolls in the direction $\phi>0$ until the end of inflation is reached at $\phi=\phi_{END}$. To ensure enough inflation $a_2$ has to be small $a_2\lesssim 1/100$. The other parameters are not necessarily small.

The derivatives of the potential are the following,
\begin{eqnarray}
\frac{V'}{V}&=&\frac{-2 a_2 \phi-3 a_3 \phi^2 -4 a_4 \phi^3 +\cdots}{1-a_2 \phi^2-a_3
\phi^3 -a_4 \phi^4 +\cdots}
\\
\frac{V''}{V}&=&\frac{-2 a_2- 6 a_3 \phi -12 a_4 \phi^2 +\cdots}{1-a_2 \phi^2-a_3
\phi^3 -a_4 \phi^4 +\cdots}
\\
\frac{V'''}{V}&=&\frac{- 6 a_3 -24 a_4 \phi +\cdots}{1-a_2 \phi^2-a_3
\phi^3 -a_4 \phi^4 +\cdots}.
\end{eqnarray}
One of the key features of the SUGRA models is that only a finite number of parameters are tuned to be small and the rest of the coefficients $a_i$ are large, i.e. of order one or larger. Since the end of inflation occurs when $V'/V=\sqrt{2}$ it is determined in these models by terms other than the quadratic. Then the value of the field at the end of inflation  is typically small $\phi_{END}\lesssim 1$ and the potential is still dominated by
the constant (the 1).  The smallness of $\phi_{END}$ is robust and is a key factor that determines the phenomenological consequences. The value of the field approximately 50 to 60 e-folds before the end of inflation
$\phi_{CMB}$, determines the CMB quantities, and can be determined either by the
quadratic term, or by the higher order terms.

\subsection{Simple potentials}

\subsubsection{A quadratic model}

We first analyze the simplest quadratic model to show that it is not a true representative of the small field models that appear as models of modular inflation. We use the discussion also to introduce notations and conventions.

If the potential is quadratic
\begin{equation}
V(\phi)= \Lambda^4 \left(1-a_2 \phi^2 \right)
\end{equation}
$\eta$ is essentially constant for the relevant range of $\phi$,
$\eta=-2 a_2$.
The number of e-folds as a function of the field is given by
\begin{equation}
N(\phi)\simeq - \frac{1}{2a_2}\ln\left(\frac{\phi_N}{\phi_{END}}\right)
\end{equation}
hence
\begin{equation}
\phi_N=\phi_{END}e^{-2a_2N}.
\end{equation}
The value of the field at the end of inflation  $\phi_{END}$ is the value of the field when $\epsilon(\phi)=1$:
$\frac{V'}{V}=\sqrt{2}$, that is when
\begin{equation}
\frac{2 a_2 \phi_{END}}{1-a_2 \phi_{END}^2}=\sqrt{2}
\end{equation}
so
\begin{equation}
a_2 \phi_{END}^2+\sqrt{2} a_2 \phi_{END}-1=0
\end{equation}
and assuming that $a_2$ is small ($2a_2=-\eta$)
\begin{equation}
\phi_{END}\simeq 1/\sqrt{a_2}=\sqrt{\frac{2}{-\eta}}.
\end{equation}
Here we immediately see that if $\phi_{END}\sim 1$ then $\eta$ has to be large and vice versa, if $\eta\sim -1/50$ then $\phi_{END}\sim 10  \frac{1}{\sqrt{|\eta|50}}$. Obviously this is not really a small field model and it is unlikely to be realized in the context of modular inflation.

We turn now to evaluate the amplitude of GW. For this we need to know the value of $\epsilon$ as a function of $N$,
\begin{equation}
\epsilon_{N}=\frac{1}{2}\left(2a_2 \phi_N\right)^2=\frac{1}{2}\left(\eta\sqrt{\frac{2}{\eta}}e^{\eta N}\right)^2
\end{equation}
(recall that $\eta$ is negative) so
\begin{equation}
\epsilon_{N}=|\eta|e^{-2|\eta| N}.
\end{equation}
The maximum possible value of $\epsilon_{N}$ is determined by $\partial_{|\eta|} \epsilon=0$ and is obtained at $\eta=-1/2N$. In this case
\begin{equation}
\epsilon_{max}=\frac{1}{2N_{CMB}}e^{-1}=3\times 10^{-3} \left( 60/N_{CMB}\right)
\end{equation}
and
\begin{equation}
r_{max}=5\times 10^{-2} \left(60/N_{CMB}\right).
\end{equation}
This is a detectable value in future $CMB$ experiments. However, in this case $\phi_{END,max}\sim 15 \sqrt{N_{CMB}/60}$ which cannot be realized in our scenario. The value of the spectral index for the parameters that give maximal $r$ is less than unity $n_S=1-1/N\simeq .98$ (for $N_{CMB}=60$).

The expression for the running of the spectral index for this model is substantially different from the expression for the more general models that we will describe later and therefore we do not discuss it in detail. The final result for the running parameter $\alpha$ is
\begin{equation}
\alpha_N=-16|\eta|^2e^{-|\eta|N},
\end{equation}
so the maximal $\alpha$ is obtained when $|\eta|=2/N$,
\begin{equation}
\alpha_{max}=-\left(\frac{2}{15 e}\right)^2\left(\frac{60}{N_{CMB}}\right)^2\simeq -2.4 \times 10^{-3}\left(\frac{60}{N_{CMB}}\right)^2.
\end{equation}
If the Planck satellite will measure $\alpha$ with an accuracy of $\text{a few} \times 10^{-3}$ this is barely detectable. The value of the spectral index for the maximal $\alpha$ is $n_S=1-4/N_{CMB}$ so $n_S=.93$ for $N_{CMB}=60$.

\subsubsection{Higher order models}

We have already seen that $p=2$ cannot be considered as a small field model and it is unlikely to be realized in the modular inflation context. The next simple case that we analyze is when the deviation of the potential from a constant in the range of field values from $\phi_{CMB}$ to $\phi_{END}$ is dominated by a single higher order term $a_p \phi^p$, $p > 2$. These models will indeed turn out to be small field models. However,  for such models to be relevant $p-2$ parameters $a_2, a_3, \cdots, a_{p-1}$ have to be tuned to be small. We therefore do not expect models with a large $p$ to be relevant as representatives of realistic modular inflation models. We can perhaps expect that some low order coefficients are set to zero by some symmetry argument or an additional tuning.  However, we do  discuss models with arbitraty $p$ to demonstrate that even with a large amount of fine tuning it is not possible to generate a large amplitude of GW and a substantial amount of spectral index running.

The potential in this case is the following
\begin{equation}
V(\phi)= \Lambda^4 \left(1-a_p \phi^p \right)
\end{equation}
and
\begin{equation}
\epsilon(\phi)=\frac{1}{2}\left(\frac{-p a_p \phi^{p-1}} {1-a_p \phi^p}\right)^2.
\end{equation}
The end of inflation is determined by
\begin{equation}
\frac{1}{\sqrt{2}}\frac{p a_p(\phi_{END})^{p-1}} {1-a_p (\phi_{END})^p}=\pm  1.
\end{equation}
Assuming that the potential is still dominated by the constant also at the end of
inflation
\begin{equation}
\label{phiENDp}
\phi_{END}=\left(\frac{\sqrt{2}}{p a_p}\right)^{1/(p-1)}.
\end{equation}
This approximation needs to be checked for specific models.

\begin{equation}
N(\phi)= \int_\phi^{\phi_{END}}\frac{d\widetilde{\phi}}{p a_p
(\widetilde{\phi})^{p-1}} - \int_\phi^{\phi_{END}}\frac{d\widetilde{\phi}}{p}
\widetilde{\phi}
\end{equation}
which can be evaluated exactly,
\begin{equation}
N(\phi)=- \frac{1}{p(p-2) a_p}\frac{1}{\widetilde{\phi}^{p-2}}{\hbox{\huge
$|$}}_{\phi}^{\phi_{END}} - \frac{1}{2p} \widetilde{\phi}^{2}{\hbox{\huge
$|$}}_{\phi}^{\phi_{END}}
\end{equation}
If $\phi\ll\phi_{END}$, and $\phi_{END}\lesssim 1$ then
\begin{equation}
N(\phi)\simeq  \frac{1}{p(p-2) a_p}\frac{1}{{\phi}^{p-2}}
\end{equation}
independently of $\phi_{END}$, and
\begin{equation}
 \label{phiNp}
\phi_N= \left( \frac{1}{p(p-2) a_p}\frac{1}{N}\right)^{\frac{1}{p-2}}
\end{equation}
so
\begin{equation}
\eta_N= -p(p-1) a_p {\phi_N}^{p-2}=-\frac{1}{N}\frac{p-1}{p-2}
\end{equation}
independently of $a_p$ or any other mass parameter in the potential.
\begin{equation}
\eta_{CMB}= -\frac{1}{N_{CMB}}\frac{p-1}{p-2}
\end{equation}
and
\begin{equation}
(n_S)_{CMB}= 1- \frac{2}{N_{CMB}}\frac{p-1}{p-2}
\end{equation}
which is the result that Kinney et al get \cite{kinney}.
\begin{equation}
\epsilon_{N}= \frac{1}{2}\left(\frac{V'(\phi_N)}{V(\phi_N)}\right)^2\simeq
\frac{1}{2}\left(p a_p (\phi_N)^{p-1}\right)^2=
 \frac{1}{2}\left(\frac{\eta_N \phi_N}{p-1}\right)^2\ll (\eta_N)^2.
\end{equation}
Let us evaluate this more precisely, using eq.(\ref{phiNp}), $\phi_N^{p-1}= \left(\frac{1}{p(p-2) a_p N} \right)^{\frac{p-1}{p-2}}$
\begin{equation}
\epsilon_{N}= \frac{1}{2}\left(\frac{1}{p a_p} \right)^{\frac{2}{p-2}} \left(\frac{1}{(p-2) N} \right)^{2\frac{p-1}{p-2}}.
\end{equation}
Using eq.(\ref{phiENDp}) we may express $\epsilon_N$ as
\begin{equation}
\epsilon_{N}= \left(\frac{\phi_{END}}{\sqrt{2}(p-2) N} \right)^{2\frac{p-1}{p-2}}.
\end{equation}

One way of understanding the implications of the results is to tabulate the value of $r$ for $\phi_{END}=1$, and $N_{CMB}=60$. This is done in Table~\ref{rsimple} where the corresponding value of $\alpha$ (calculated below) is also listed.
\begin{table}
\begin{center}
\begin{tabular}{|l|c|c|c|c|c|c|c|}
  \hline
   \ $p$ & 3 & 4 & 5 &  7 & 10 & $p\to\infty $ \\
    \hline
   \ $r$ & $3.1 \times 10^{-7}$ & $3.3 \times 10^{-6}$ & $6.1 \times 10^{-6}$ & $ 7.9\times 10^{-6}$ & $6.8\times 10^{-6}$ & 0 \\
   \hline
    \ $\alpha$ & $5.6 \times 10^{-4}$ & $4.2 \times 10^{-4}$ & $3.7 \times 10^{-4}$ & $3.4 \times 10^{-4}$ & $3.4 \times 10^{-4} $ & $2.8 \times 10^{-4}$ \\
    \hline
\end{tabular}
\end{center}
\caption{The values of $r$ and  $\alpha$ for simple models assuming that $\phi_{END}=1$ and $N_{CMB}=60$.}
\label{rsimple}
\end{table}

Alternatively we can look for the largest possible $r$ by maximizing $\epsilon$.
\begin{equation}
\frac{\partial\epsilon_N}{\partial p}= -2 \left(\frac{\phi_{END}}{\sqrt{2}(p-2) N} \right)^{2\frac{p-1}{p-2}} \frac{1}{(p-2)^2}\left((p-1)+\ln\left( \frac{\phi_{END}}{\sqrt{2}(p-2) N}\right)\right),
\end{equation}
so $\epsilon_N$ is maximized for
\begin{equation}
\phi_{END}=\sqrt{2}N(p-2)e^{-(p-1)}.
\end{equation}
For this specific value of $\phi_{END}$, $r$ will be maximal for a specific value of $p$. For example for $\phi_{END}=1$ the maximal value of $r$ is obtained for $p$ is about $7$ as can be seen in Table~\ref{rsimple}.
Then
\begin{equation}
\epsilon_{N,max}=e^{-2\frac{(p-1)^2}{p-2}}.
\end{equation}
The values for $r$ and $\phi_{END}$ are listed  in Table~\ref{rsimplephi}, where the corresponding values of $\alpha$ are also listed.
\begin{table}
\begin{center}
\begin{tabular}{|c|c|c|c|c|c|c|c|} \hline
    \ $p$ & $3$ & $4 $& $5$ &  $7$ & $10$ & $p\to\infty$ \\ \hline
    \ $r$ &$5.3\times 10^{-3}$ &$1.9 \times 10^{-3}$ &$3.7 \times 10^{-4}$ & $ 8.9\times 10^{-6}$ &$ 2.6\times 10^{-8}$ & $0$\\ \hline
    \ $\alpha$ & $5.6 \times 10^{-4}$ & $4.2 \times 10^{-4}$ & $3.7 \times 10^{-4}$ & $3.4 \times 10^{-4}$ & $3.4 \times 10^{-4} $ & $2.8 \times 10^{-4}$ \\
    \hline
    $\phi_{END}$ &$11.5$ &$8.4$  &$4.7$ &  $1.1$ & $0.08$& $0$ \\ \hline
 \end{tabular}
\end{center}
\caption{ The dependence of $r$ and $\alpha$ for simple models on $\phi_{END}$ for $N_{CMB}=60$. In the approximation we use $\alpha$ does not depend on $\phi_{END}$.}
\label{rsimplephi}
\end{table}

Both Tables~\ref{rsimple} and \ref{rsimplephi}  show that $r$  and $\alpha$ are too small and unobservable for these models when $\phi_{END}\lesssim 1$.

For this class of models the spectral index running parameter can be approximated as $\alpha= - 16 \epsilon \eta + 24 \epsilon^2 + 2\xi^2\simeq 2 \xi^2$ because $\epsilon$ is very small.
Then
\begin{equation}
\alpha\simeq 2 (p-1)(p-2) (p a_p \phi^{(p-2)})^2.
\end{equation}
Using eq.(\ref{phiNp}) we find
\begin{equation}
\alpha_N= 2 \frac{(p-1)}{(p-2)} \frac{1}{N^2} 
\end{equation}
so
\begin{equation}
\alpha_{CMB} = 2.8 \times 10^{-4} \frac{(p-1)}{(p-2)} \left(\frac{60}{N_{CMB}}\right)^2
\end{equation}

\subsection{More complicated models}

We wish to consider more complicated models to find out whether it is possible by some additional tuning to find models that do produce a substantially larger amount of GW or spectral index running. We find that an enhancement of the GW amplitude is possible if the potentials are extremely fine tuned. However, even with massive fine tuning it is not possible to make the GW amplitude observable. We find that the parameter $\alpha$ can not be increased significantly for this class of models.

The more complicated potentials that we wish to consider are those for which the
end of inflation is determined by some higher order term and the region near
$\phi_{CMB}$ is determined by the quadratic term. Other
combinations of terms can, of course, be considered. The reason that we focus our attention on such models is
that they seem to have a better chance of producing very small but still observable gravity
wave component since in this cases $a_2 \phi_{CMB}$ can be made larger than the higher order term.

We will consider the following potentials
\begin{equation}
V(\phi)= \Lambda^4 \left(1-a_2 \phi^2 -a_p \phi^p \right).
\end{equation}
Obviously, a model with large $p$ requires that $p-3$ parameters are tuned to be small. Therefore we do not expect models with a large $p$ to be realistic. Again, we consider them for the purpose of showing that even with an extensive fine tuning it is not possible to produce an observable amount of GW or running parameter $\alpha$.
As in the previous subsection, $\alpha$ can be approximated as $\alpha\simeq 2 \xi^2$.

As a warmup example let us consider a potential that has a quadratic and cubic term $p=3$.
We consider the case that  $a_2 \ll a_3$, such as expected when the second
derivative is tuned to be small but the third derivative does not need to be
particularly small. In this case
\begin{equation}
\label{epsp3}
\epsilon(\phi)=\frac{1}{2}\left(\frac{- 2 a_2 \phi- 3 a_3 \phi^{2}} {1- a_2
\phi^2-a_3 \phi^3}\right)^2
\end{equation}
and
\begin{equation}
N(\phi)= \int_\phi^{\phi_{END}} d\widetilde{\phi}\ \frac{1- a_2 \phi^2-a_3
\phi^3}{2 a_2 \phi+ 3 a_3 \phi^2}
\end{equation}
which can be evaluated exactly,
\begin{eqnarray}
N(\phi)&=&\Biggl[-\frac{1}{18 a_3} \widetilde{\phi}( 2 a_2 + 3 a_3 \widetilde{\phi})
+\frac{1}{2 a_2} \ln\left[\widetilde{\phi} \right]-\frac{1}{2 a_2} \ln\left[2 a_2 +
3 a_3 \widetilde{\phi} \right] \cr &+& \frac{2}{27} \left(\frac{a_2}{a_3}\right)^2
\ln\left[2 a_2 + 3 a_3 \widetilde{\phi} \right] \Biggr]_{\phi}^{\phi_{END}}.
\end{eqnarray}
Since $\frac{a_2}{a_3}\ll 1$ we may approximate $N(\phi)$ by
\begin{equation}
N(\phi)=\Biggl[-\frac{1}{18 a_3} \widetilde{\phi}( 2 a_2 + 3 a_3 \widetilde{\phi})
+\frac{1}{2 a_2} \ln\left[\frac{\widetilde{\phi} }{2 a_2 + 3 a_3
\widetilde{\phi}}\right] \Biggr]_{\phi}^{\phi_{END}}.
\end{equation}

Several parameter regions exist. If  $3 a_3 \phi_{CMB}>
2 a_2$ then also $  3 a_3 \phi_{END}> 2 a_2  $. In this case the correct approximation is of course just to take the cubic term, and then this case is similar to the previous simple $p=3$ model. If $2 a_2 > 3 a_3 \phi_{END}$, then it is also true that $2 a_2 > 3 a3 \phi_{CMB}$, and the end of inflation will be determined by the quadratic term, going back to the simple quadratic case.  So the only interesting case is when  $2 a_2> 3 a_3 \phi_{CMB}$, and $ 3 a_3 \phi_{END}> 2 a_2  $, the quadratic term determines the evolution near $\phi_{CMB}$ and the end of inflation is determined by the cubic term (in general some higher order terms). In this interesting case, taking into account that $\phi_{CMB}< \phi_{END}$ we find,
\begin{equation}
N(\phi)=-\frac{1}{6} \phi_{END}^2 + \frac{1}{2 a_2} \ln\left[\frac{1 }{ 3
a_3}\right] - \frac{1}{2 a_2} \ln\left[\frac{\phi} {2 a_2}\right]= -\frac{1}{6}
\phi_{END}^2 - \frac{1}{2 a_2} \ln\left[\frac{3 a_3 } {2 a_2}\phi\right].
\end{equation}
In this case we may also estimate $\phi_{END}$, using eq.(\ref{epsp3}) and assuming that the constant still dominates the potential towards the end of inflation,
\begin{equation}
3 a_3 \phi_{END}^2 =\sqrt{2}
\end{equation}
so
\begin{equation}
\label{phiENDp3}
\phi_{END} =\sqrt{\frac{\sqrt{2}}{3 a_3 }}.
\end{equation}
We may compare $a_3 \phi_{END}^3= 2^{3/4} \sqrt{\frac{1}{3 a_3 }}$ to 1, and see
that if $a_3 >1$ then $\phi_{END}<1$ and the constant indeed dominates.

In this case we may further approximate $N(\phi)$
\begin{equation}
N(\phi)= - \frac{1}{2 a_2} \ln\left[\frac{3 a_3 } {2 a_2}\phi\right]
\end{equation}
which can be inverted,
\begin{equation}
\phi_N = \frac{2 a_2}{ 3 a_3} e^{- 2 a_2 N}.
\end{equation}
In particular,
\begin{equation}
\phi_{CMB} = \frac{2 a_2}{ 3 a_3} e^{- 2 a_2 N_{CMB}}.
\end{equation}
Since in this case
\begin{equation}
\eta_{CMB} \simeq -2 a_2
\end{equation}
then
\begin{equation}
\label{phi32}
\phi_{CMB} = \frac{|\eta_{CMB}|}{ 3 a_3} e^{- |\eta_{CMB}| \cdot N_{CMB}}
\end{equation}
and
\begin{equation}
\epsilon_{CMB} \simeq \frac{1}{2} (2 a_2 \phi_{CMB} )^2 = \frac{1}{2} \frac{1}{9
a_3^2} (\eta_{CMB})^4
 e^{-2|\eta_{CMB}| N_{CMB}}.
\end{equation}
Using eq.(\ref{phiENDp3}) we express $\epsilon_{CMB}$ in terms of $\phi_{END}$,
\begin{equation}
\epsilon_{CMB} \simeq \frac{1}{4}(\phi_{END})^4  (\eta_{CMB})^4
 e^{-2|\eta_{CMB}| N_{CMB}}
\end{equation}
We may now check whether or not  $2 a_2 > 3 a_3 \phi_{CMB}$ as we have assumed. Equivalently we check whether $|\eta_{CMB}| > 3 a_3 \phi_{CMB}$. Since
\begin{equation}
3 a_3 \phi_{CMB} = \eta_{CMB}  e^{- |\eta_{CMB}|N_{CMB}}.
\end{equation}
As we can see, $2 a_2 > 3 a_3 \phi_{CMB}$ is generically true if $|\eta_{CMB}| \gtrsim 1/N_{CMB}$.

For this case
\begin{equation}
n_S =1- 2|\eta_{CMB}|\lesssim 1-1/30=.97
\end{equation}
and
\begin{equation}
r= 16 \epsilon = 4 (\phi_{END})^4  (\eta_{CMB})^4
 e^{-2|\eta_{CMB}| N_{CMB}}.
\end{equation}
The maximal $r$ for fixed $\phi_{END}$ and $N_{CMB}$ is reached at $|\eta_{CMB}|=2/N_{CMB}$ when
\begin{equation}
r_{max}= 4 (\phi_{END})^4  \left(\frac{2}{N}\right)^4\
 e^{-4}=9\times 10^{-8} \left( 60/N_{CMB}\right)^4(\phi_{END})^4,
\end{equation}
somewhat smaller than the pure cubic case as in Table~\ref{rsimple}. We see that the attempt to enhance the GW amplitude by an additional tuning has not succeeded for the case $p=3$. As we will show an enhancement can be obtained for $p>3$.

The running spectral index parameter $\alpha$ is approximately given by
\begin{equation}
\alpha= 48 a_2   a_3 \phi.
\end{equation}
Using eq.(\ref{phi32})  we may express $\alpha_{CMB}$ as
\begin{equation}
\alpha_{CMB}= 4 |\eta_{CMB}|^2 e^{-|\eta_{CMB}| N_{CMB}}.
\end{equation}
The maximal value of $\alpha_{CMB}$ is obtained for $\eta_{CMB}=-2/N_{CMB}$,
\begin{equation}
\alpha_{max}= \frac{16}{e^2 N_{CMB}^2}= 6\times 10^{-4}\left( 60/N_{CMB}\right)^2 .
\end{equation}

The more general case for arbitrary $p$ is presented below.
\begin{equation}
V(\phi)= \Lambda^4 \left(1-a_2 \phi^2 -a_p \phi^p \right)
\end{equation}
\begin{eqnarray}
N(\phi)&=& \int\frac{1-a_2\phi^2-a_p\phi_p^p}{2a_2\phi+pa_p\phi_p^{p-1}} \\
& \simeq& \frac{1}{2 a_2} \ln\left[\frac{\phi}{\left(2a_2+pa_p\phi^{p-2}\right)^{\frac{1}{p-2}}}\right] \Biggl|^{\phi_{END}}_{\phi_N} \cr
&\simeq& \frac{1}{2 a_2}\left( \ln\left[\frac{1}{\left(pa_p\right)^{\frac{1}{p-2}}}\right]- \ln\left[\frac{\phi_N}{\left(2a_2\right)^{\frac{1}{p-2}}}\right]\right) \\
&=& -\frac{1}{2 a_2} \ln\left[\phi_N\left(\frac{pa_p}{2a_2}\right)^{\frac{1}{p-2}}\right]
\end{eqnarray}
and
\begin{equation}
\phi_N=\left(\frac{2a_2}{pa_p}\right)^{\frac{1}{p-2}} e^{-2a_2 N}.
\end{equation}
We may express $\phi_N$ in terms of $\phi_{END}$ using eq.(\ref{phiENDp}) which is also relevant to this case,
\begin{equation}
\label{phincomp}
\phi_N=\left(\phi_{END}\right)^{\frac{p-1}{p-2}}\left(\frac{2a_2}{\sqrt{2}}\right)^{\frac{1}{p-2}} e^{-2a_2 N}.
\end{equation}
In this case
\begin{eqnarray}
\epsilon_N &=&\frac{1}{2} \left(2a_2\phi_N\right)^2 \\
& =&
\left(\phi_{END}\right)^{2 \frac{p-1}{p-2}}\left(\frac{2a_2}{\sqrt{2}}\right)^{2 \frac{p-1}{p-2}} e^{-4a_2 N}
\end{eqnarray}
The maximal $\epsilon_N$ as a function of $|\eta_{CMB}|\sim 2 a_2$ for fixed $\phi_{END}$, and $N$ is obtained at
\begin{equation}
2a_2=\frac{1}{N}\frac{p-1}{p-2}
\end{equation}
and the maximal value is
\begin{equation}
\epsilon_{N,max}=\left[\left(\frac{\phi_{END}}{e\sqrt{2} }\right)\frac{1}{N}{ \frac{p-1}{p-2}}\right]^{2 \frac{p-1}{p-2}}\
\end{equation}
so
\begin{equation}
r_{max}=16\left[\left(\frac{1}{60e\sqrt{2} }\right){ \frac{p-1}{p-2}}\right]^{2 \frac{p-1}{p-2}} \left(\phi_{END}\right)^{2 \frac{p-1}{p-2}} \left(\frac{60}{N_{CMB}}\right)^{2 \frac{p-1}{p-2}}.
\end{equation}
The maximal values of $r$ and the corresponding values of $\alpha$ are listed in Table~\ref{rcomplicated},
\begin{table}
\begin{center}
\begin{tabular}{|c|c|c|c|c|c|c|}
  \hline
    \ $p$ & $3$ & $4$ & $5$ & $7 $& $10$ &$ p\to\infty $\\
    \hline
    \ $r_{max} $&\ $ 9.0 \times 10^{-8}$\ &\ $4.4 \times 10^{-6}$\ &\ $1.7 \times 10^{-5}$\ &\ $5.3\times 10^{-5}$\ &\ $1.0\times 10^{-4}$\ &\ $ 3.0\times 10^{-4}$\ \\
     \hline
     \ $\alpha$ & $6.0 \times 10^{-4}$ & $3.7 \times 10^{-4}$ & $2.1 \times 10^{-4}$ & $6.0 \times 10^{-5}$ & $6.2 \times 10^{-6}$ & 0 \\
    \hline
    \hline
     \ $\alpha_{max}$ & $6.0 \times 10^{-4}$ & $4.5 \times 10^{-4}$ & $4.0 \times 10^{-4}$ & $3.6 \times 10^{-4}$ & $3.4 \times 10^{-4}$ & $3.0\times 10^{-4}$ \\
     \hline
     \ $r $&\ $ 9.0 \times 10^{-8}$\ &\ $3.5 \times 10^{-6}$\ &\ $1.0 \times 10^{-5}$\ &\ $1.9\times 10^{-5}$\ &\ $2.0\times 10^{-5}$\ &\ $ 0$\ \\
   \hline
 \end{tabular}
\end{center}
\caption{ Listed in the first two lines are the maximal value of $r$ in the CMB and the corresponding value of $\alpha$ assuming that $\phi_{END}=1$. In the last two lines the maximal value of $\alpha$ and the corresponding value of $r$ are listed. All values are for the more complicated models discussed in the text.}
\label{rcomplicated}
\end{table}
Comparing Table~\ref{rcomplicated} to Table~\ref{rsimple} it is possible to see that the amount of GW for the highly fine tuned models with a large $p$ is much larger than the simple monomial case. Still, all values are way below delectability for $\phi_{END}\lesssim 1$. A detectable signal will require $\phi_{END}\gtrsim 10$

For the models we consider $\alpha$ is approximately given by
\begin{equation}
\label{alphacomp}
\alpha=4a_2p(p-1)(p-2)a_p\phi^{p-2}.
\end{equation}
Substituting eq.(\ref{phincomp}) into eq.(\ref{alphacomp})  we get,
\begin{equation}
\alpha_N=2(p-1)(p-2)(2a_2)^2e^{-2a_2N(p-2)}.
\end{equation}
Approximating $|\eta|=2a_2$, the maximal running is when $\eta \simeq \frac{2}{N(p-2)}$.
Then
\begin{equation}
\alpha_{N,max}\simeq \frac{8(p-1)}{ e^2 (p-2) N^2}
\end{equation}
\begin{equation}
\alpha_{max}= 3 \times 10^{-4}\ \frac{p-1}{p-2} \left(\frac{60}{N_{CMB}}\right)^2
\end{equation}
which is again undetectable.

The values of $\eta$ for which $r$ and $\alpha$ are maximized are different in this case.

\section{Summary and conclusions}
\label{conc}

In this paper we have discussed small field models of modular inflation.
We have explained the difficulties in realizing them as low energy SUGRA models of the form expected as the low energy effective action of string theory. We have found that the difficulties were associated with the specific nature of the geometry of the moduli space in the perturbative region. We then discussed SUGRA models with a canonical K\"ahler potential as examples for models that can be expected in regions where the perturbative geometry is substantially modified. We have determined the constraints that they have to satisfy to be viable inflation models. We believe that the likely region where such models can be realized is  the central region of moduli space where the coupling and compact volume are of order unity.

The successful SUGRA models generically predict a very small GW amplitude $r \lesssim 10^{-4}$ for $\phi_{END}\leq 1$ and a very small running of the spectral index $\alpha<10^{-3}$. This means that a detection of GW in the CMB or spectral index running in the foreseeable future will rule out these models.

Our results explain in retrospect many of the past difficulties encountered in attempting to find good modular inflation models.
More generally, the condition that we have found on the curvature of field space in SUGRA has to be met by any successful small field SUGRA model of inflation.

\section*{Acknowledgements}
The research of IBD and RB is supported in part by ISF grant 470/06.
The research of SdA is supported in part by the US Department of Energy under grant DE-FG02-91-ER-40672.
IBD would like to thank the Hebrew University in Jerusalem for their hospitality during the duration of the research.

\end{document}